# Increasing risk behaviour can outweigh the benefits of antiretroviral drug treatment on the HIV incidence among men-having-sex-with-men in Amsterdam

Shan Mei[*,1,2], Rick Quax[2] , David van de Vijver[3] , Yifan Zhu[1] , A.V. Boukhanovsky [4] and P.M.A. Sloot[2,4]

[1] Information System and Management College, National University of Defense Technology, Changsha, P. R. China

[2] Computational Science, University of Amsterdam, Amsterdam, Netherlands

[3] Dept. of Virology, Erasmus MC, University Medical Centre Rotterdam, Netherlands

[4] ITMO National Research University, St. Petersburg, Russian Federation

Email: Shan Mei[*]- MeiShan.ann@Gmail.com; Rick Quax - r.quax@uva.nl; David van de Vijver - d.vandevijver@erasmusmc.nl; Yifan Zhu - yfzhu@nudt.edu.cn; P.M.A. Sloot - P.M.A.Sloot@uva.nl;

[*]Correspondingauthor

# Abstract

Background: The transmission through contacts among MSM (men who have sex with men) is one of the dominating contributors to HIV prevalence in industrialized countries. In Amsterdam, the capital of the Netherlands, the MSM risk group has been traced for decades. This has motivated studies which provide detailed information about MSM's risk behavior statistically, psychologically and sociologically. Despite the era
of potent antiretroviral therapy, the incidence of HIV among MSM increases. In the long term the contradictory effects of risk behavior and effective therapy are still poorly understood.

Methods: Using a previously presented Complex Agent Network model, we describe steady and casual partnerships to predict the HIV spreading among MSM. Behavior-related parameters and values, inferred from studies on Amsterdam MSM, are fed into the model; we validate the model using historical yearly incidence data. Subsequently, we study scenarios to assess the contradictory effects of risk behavior and effective therapy, by varying corresponding values of parameters. Finally, we conduct quantitative analysis based on the resulting incidence data.

Results: The simulated incidence reproduces the ACS historical incidence well and helps to predict the HIV epidemic among MSM in Amsterdam. Our results show that in the long run the positive influence of effective therapy can be outweighed by an increase in risk behavior of at least 30% for MSM.

Conclusion: We recommend, based on the model predictions, that lowering risk behavior is the prominent control mechanism of HIV incidence even in the presence of effective therapy.

# Background

The HIV epidemic is a global challenge and has a destructive impact on human life and welfare. One of the major or even dominating contributors to HIV prevalence in industrialized countries is the transmission through contacts among men who have sex with men (MSM). Particularly, in the Netherlands, MSM still accounts for the highest incidence of Sexually Transmitted Infections (STIs) including HIV [1]. In fact, the proportion of MSM in new HIV cases recorded at the HIV registry continued to increase from 44% in 2003 to 59% in 2006 [1, 2].

Amsterdam harbors the majority of the largest MSM risk group in the Netherlands that has been traced and recorded for decades, providing us with detailed information about Amsterdam MSM's behavior characteristics. The Amsterdam Cohort Study (ACS) is such a study which has been conducted since 1984 with thousands of MSM recruited.

In order to embody heterogeneous behavior such as personal infection progressions and steady/casual partnership linkages, we use agent-based modeling to simulate the HIV spreading among MSM in Amsterdam, based on a previously presented Complex Agent Network model [3]. This model helps to describe partnerships in detail, with (steady or casual) partnerships, durations and frequencies of sexual actions, influencing the HIV spreading among MSM in a well-structured *in silico* population. This enables modeling realistically heterogeneous populations and incorporation of social behavior data in the model. Risk behavior is of vital significance to HIV transmission and has sparked many psychological surveys. Questionnaires completed by 324 MSM aged 18-34 were conducted by Davidovich et al., to assess sexual behavior and related cognitions [4]. The results reveal, amongst others, that unprotected anal intercourse (UAI) is regarded as a symbol of trust and believing that the partner desires UAI is associated with less condom use; additionally, perceiving UAI as more gratifying was associated with having riskier UAI. Unfortunately, beliefs of people living with HIV about their own responsibility for preventing HIV transmission (personal responsibility) and their sex partners' responsibility for protecting themselves (partner responsibility) are poorly understood [5].

Despite psychological and sociological reasons, the so-called "safer sex fatigue" and excessive confidence in low viral load further intensify the reluctance on reducing risky behavior. Crepaz et al., concluded [6] that people's beliefs about Highly Active Antiretroviral Therapies (HAART) and viral load may promote unprotected sex, although HIV-positive patients receiving HAART did not exhibit increased sexual risk behavior, even when therapy achieved an undetectable viral load. In Amsterdam, the worrying increase of STIs was investigated in studies carried out by the department of AIDS research of the municipal health service. The result indicated that the introduction of HAART may have had an influence on the increase of STIs and risky sexual behavior [7].

Increase in risk behavior and resurgent epidemics have been reported post-HAART, and resurgent epidemics is most likely caused by increasing sexual risk behavior [8]. Furthermore, low diagnosis fraction of infections may intensify this threat, since HIV-positive men who are unaware of their serostatus have higher risk behavior than the aware group [9–11]. Regarding the ACS, only 40% MSM accepted HIV testing at their STI consultation, among the recently infected at the STI clinic [12].

In this study, the primary objective is to explain and reproduce historical data and then assess the potential influence of varying risk behavior and therapy factors, with the aim to understand and predict the HIV spreading among MSM in Amsterdam. For this we need to combine sexual contact networks with HIV infection dynamics and individual behavior. To the best of our knowledge this is the first report on this combined evidence-based computational epidemics model.

## Methods
Modeling the HIV epidemic in Complex Agent Networks

Using a previously developed Complex Agent Network [3], we model the HIV epidemics among MSM in Amsterdam through a sexual contact network in which each node represents an individual and edges represent potentially infectious contacts. Please see Fig. 2 in [13] as a schematic illustration of virus spreading in a network. The number of edges emanating from a node is the *degree* of the node and the *degree distribution* is the distribution of this quantity across the population [14]. In this study, the degree of a node, drawn from a specific degree distribution, is assumed to be static while the neighboring nodes that this node links to may change over time. There are classes of network geometry such as small-world, Poisson and scale-free networks. Since scale-free networks have been shown to be the representative of a variety of human sexual contact networks [15–17], we perform stochastic simulations of the spreading of HIV in undirected scale-free networks. The scale-free networks are characterized by a highly skewed distribution of contacts so that most of the nodes are weakly connected and a small number of nodes have very high connectivity. In particular, the degree distribution of this class of networks follows a power law distribution, i.e., $p_k = Ak^{-\gamma}$ where $A$ is a normalization factor. After each node is assigned with a degree drawn from a given power law, these nodes are connected at random according to the so-called configuration model [18,19] with the guarantee that the sum of all degrees is an even number and that there are no loops (nodes connected to themselves) and multiple edges (between two nodes).

The disease status of infected individuals is heterogenous with respects to infection stages, disease durations and infectiousness, etc., so that we can model individuals as agents and individuals' status transitions as internal changes of agents. The HIV infection progression occurs in roughly three stages: primary infection (PI), asymptomatic period (AP) and AIDS. Regarding status transition, for instance, infection reaches from PI to AP with a delay of about 3 months. We assume an irreversible "negative → PI → AP → AIDS" transition order for each infection, excluding the case that an individual with AIDS can with the right treatment go back to the asymptomatic phase, because in general patients are most likely diagnosed and treated before reaching the AIDS stage. During the HIV epidemic, each individual's infection status is related to temporally varying infectiousness due to the corresponding different plasma HIV RNA levels. The transmission probability across each edge depends on the status of the two connected individuals. Therefore, the probability with which a susceptible individual becomes infected is determined by the connectivity (degree) of the individual and the infection status of his partners (neighboring nodes) in the network. For simplicity, we assume no HIV mutations during the course of HIV epidemic. Additionally, individuals who die of AIDS will be replaced by healthy individuals, keeping the total of MSM under study static throughout simulations.

Our model distinguishes between steady and casual relationships as two types of sexual relationships, which brings different transmission probabilities that occur across likewise steady or casual edges. Results from previous models show that the transmission dynamics of HIV differ significantly depending on relationship type, i.e., long steady relationships or short casual partnerships [20]. Moreover, there are strong indications that a substantial proportion of new HIV infections occur within steady partnerships [20,21]. Multiple sexual actions can take place along a steady edge (within a steady relationship) yearly, while only one sexual action take place along a casual edge (within a casual partnership) yearly. Each individual can have at most one steady partner with whom he stays for some time. Casual partnerships can be formed between any two individuals, but men with a steady partner have fewer partners than single men. Even steady partners may be involved in casual partnerships during or between steady relationships, though they may negotiate safety agreements to be monogamous or to have no UAI outside the relationship. Considering two partners with one infected and the other susceptible, the number of sexual acts that they have within a specified time duration, together with the infected one's infectiousness and therapy effect, the susceptible one's susceptibility, and risk behavior involving the two, determines the transmission probability from the infected one to the susceptible one across their partnership edge (see Section 3.3 in [3] for details).

Estimating parameters

Based on the Amsterdam cohort study (ACS) and literature, we estimate the sociological, virological and medical values and distributions of important parameters, and these are as listed in Table 1. The ACS of HIV infection and AIDS among MSM living in Amsterdam was initiated in 1984 and has involved 2299 men until 2006 [22]. Of the 2299 MSM, 571 were HIV-positive at study entry and 192 seroconverted during follow-up until the end of 2006. Consequently, characteristics of MSM's behavior and surveyed yearly HIV incidence over calendar years for the population are reported periodically [23]. How parameters in Table 1 are fed into the model and affect the infection probability between a HIV-positive and a HIV-negative are explicitly explained in Sec. 3.3 and 5.2.2 in [3]. Specifically, the infection probability per sexual action is computed by multiplying a multi-year statistic of transmission probability per action and several coefficients indicating, e.g., personal infectivity reduction due to treatment, personal infectivity increases due to risky behavior, and reduction of transmission probability between two steady partners if they reach a safety agreement on having no contacts outside the relationship.

Designing scenarios for predicting the HIV epidemic

The model can be used to predict the HIV epidemic by tuning the parameters related to two important factors: therapy effectiveness and risk behavior. Ref. [3] provides the value for the risk behavior factor in 1985-2005 among people infected and people not infected with HIV. It is strongly believed that risk behavior decreased until 1996 because of fear of the harm of HIV. The introdction of HAART in 1996 was associated with a slow increase in risk behavior. To investigate the counterbalancing impact of therapy effectiveness and risk behavior, we design several scenarios as listed in Table 2 to tentatively predict the HIV epidemic among a fixed number of 2299 individuals which is the total of MSM involved in ACS substudies since 1984. A reference scenario uses default values of parameters (Table 1) which remain unchanged during the course of simulations. Prediction scenarios 1-5 are designed based on various combinations of therapy reduction and risk behavior factors. We perform stochastic simulations of these scenarios from 1985 to 2044 (initialized according to the situation in 1984) with 1-year time steps and compare the results of the prediction scenarios to that of the reference scenario, with the aim to evaluate the significance of the factors. Each simulation result is an average over 30 realizations.

Results

Reference scenario

The incidence resulting from the reference scenario (RS in Table 2) simulation (1985-2020) versus the ACS historical incidence data (1985-2006) is shown in Figure 1. The curve of the simulated incidence is fluctuating less intensively compared to that of the ACS historical data. This is most likely due to the fact that the ACS has traces and records in incomprehensive populations every year (e.g., 2 HIV positives out of 511.12 in 1988 and 5 out of 374.54 in 2002, see Table 5 in [23]) while we model the HIV epidemic in a population with a fixed size of 2299.

Compared to 1.66 in 2006, the RS incidence does not change noticeably at the time steps following 2006. For example, the incidence increases to 1.69 in 2015 and 1.74 in 2020, and then declining to 1.63 in 2030

Table 1: The values or distributions of parameters (partly) for modeling the HIV epidemic.

| Parameter | Value | Description | Source |
|---|---|---|---|
| Duration of steady relationships | $DU(1, 2)^1$ | The duration of steady partnerships among Amsterdam MSM is reported to have an expected value of 1.5 years. | [20] |
| Duration of the asymptomatic stage for the untreated | $B(26, 0.5)^2$ | This stage lasts 13 years for people with a failed treatment or without any therapy. | [24, 25] |
| Duration of the asymptomatic stage for the successfully treated | $B(52, 0.5)$ | This stage lasts a mean value of 22.5 years for people with a successful treatment. And the duration is likely increased to a mean value of 26 years, thanks to the wide use of HAART and the improvement in the therapy regimens. | [26–28] |
| Frequency of sexual actions per year between steady partners (with infected at stage AP) | $P(30)^3$ | The frequency of either URAI or UIAI between steady partners is 15 per year. | [20] |
| Frequency of sexual actions between steady partners in the first 3 months and the last 9 months, respectively (with infected at stage PI) | $P(8)$ & $P(22)$ | The PI stage lasts for about 3 months which is shorter than the 1-year time step, so that we divide individuals' first year of infection into two periods. Thus, the frequencies in these two periods adds up to 30. | [20] |
| Transmission probability per URAI/UIAI[4] act (with infected at stage PI) | 0.22/0.044 | | [29–31] |
| Transmission probability per URAI/UIAI act (with infected at stage AP) | 0.011/0.0022 | The infected in the last 9 months of PI contribute the same to transmission possibility per act as they do at stage AP. | [29–31] |
| Reduction in risky behavior along casual partnerships for men who have a steady partner | 0.84 | Men may make an agreement with his steady partner to be monogamous or to have no UAI outside the relationship, leading to less risky behavior. | see Sec. 2 in [20] |
| Moderate (default) treatment-induced infectivity reduction factor[5] | $CU(0.1, 0.5)^6$ | ART can moderately reduce transmission probability by 50-90%. | [32] |
| Optimistic treatment-induced infectivity reduction factor | $CU(0.01, 0.1)$ | ART can optimistically reduce transmission probability by 90-100%. | [32] |
| Initial population size | 2299 | People involved in all substudies add up to 2299. | [22] |
| The power-law degree distribution's exponent $\gamma$ in this study | 1.6 | MSM population follows a power-law degree distribution with a value of $\gamma$ in the interval between 1.5 and 2. | [17] |
| The power-law degree distribution's maximum degree $k_{max}$ | 200 | We assume so. | |
| The fraction of vertices with a degree of 0 | 0.01 | We assume that a small portion in a population not having any contact. | |

[1] $DU(val1, val2)$ is a discrete uniform distribution with $val1$ and $val2$ as the lower and upper bounds.
[2] $B(n, p)$ is a binomial distribution.
[3] $P(\lambda)$ is a Poisson distribution.
[4] URAI, unprotected receptive anal intercourse; UIAI, unprotected insertive anal intercourse.
[5] Those who are receiving treatment can obtain an reduction in infectivity and thus decrease transmission probability per act. The untreated's infectivity remains on baseline quantities given in [29–31].
[6] $CU(val1, val2)$ is a continuous uniform distribution with $val1$ and $val2$ as the lower and upper bounds.

Table 2: Scenarios for simulating the HIV epidemic among MSM in Amsterdam (factors changed since 2006). The values of parameters change from 2006 onward in the simulations of scenarios. The risk behavior factor is a multiplicative factor of the transmission probability between two partners. For prediction scenarios, the value of the risk behavior factor is calculated by scaling the value in 2000-2005 (1.30 for both the HIV negative and positive, see Table 2 in [3]) to account for the increase in risk behavior after 2006. For example, an increase of 5% in risk behavior is assumed after 2006 for P1. The distributions of the moderate and optimistic treatment-induced infectivity reduction factors are listed in Table 1. All scenarios take a value of moderate treatment-induced infectivity reduction from 1985 to 2006.

| Scenario | Risk behavior factor | Treatment-induced infectivity reduction factor |
|---|---|---|
| Reference (RS) | 1.30 | Moderate |
| Prediction 1 (P1) | 1.05*1.30 | Moderate |
| Prediction 2 (P2) | 1.05*1.30 | Optimistic |
| Prediction 3 (P3) | 1.10*1.30 | Optimistic |
| Prediction 4 (P4) | 1.20*1.30 | Optimistic |
| Prediction 5 (P5) | 1.30*1.30 | Optimistic |

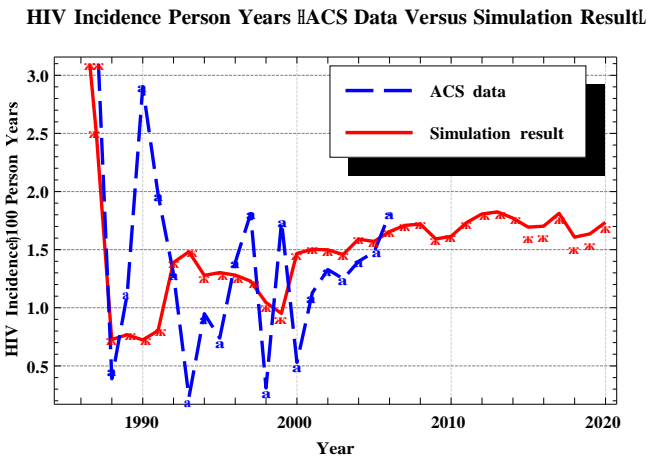

Figure 1: The incidence resulting from the RS simulation (solid) versus the ACS historical incidence data (dashed). The RS is simulated using default parameter values (Table 1) that remain unchanged throughout.

and 1.60 in 2040 (as shown in Figure 2b).

Model validation

To validate our model, a paired-sample T-test is computed, comparing the simulated incidence of RS and the ACS historical incidence in 1985-2006. The null-hypothesis under consideration is that there is no difference between these two series of data with a significance level of 0.05 (data not shown). Meanwhile, we believe that the resulting validity also applies to the testing for all the prediction scenarios that share identical parameter setting to that of RS in the range of years of 1985-2006 (see Figure 2a).

Prediction scenarios

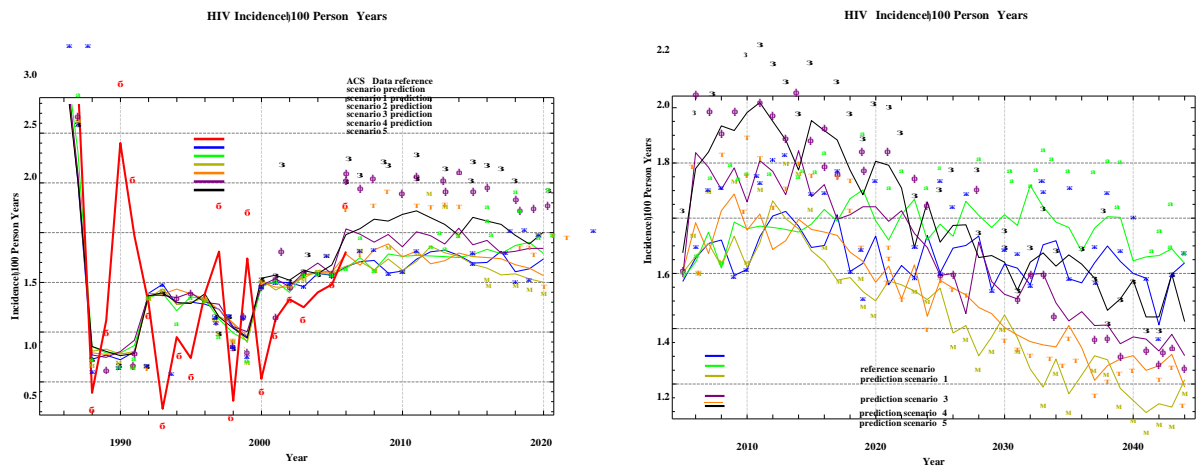

(a) The simulated incidence of the scenarios (1985-2020) and the ACS historical incidence (1985-2006)

(b) The simulated incidence of the scenarios (2005-2044)

Figure 2: Simulated incidence resulting from scenarios.

The incidence resulting from prediction scenarios, together with the RS incidence and ACS historical incidence, is shown in Figure 2. To further compare these data and assess the effect of parameter tuning, we list the simulated incidence and the corresponding increase or decline percentage in each specified year for predicting scenarios compared to the RS incidence in Table 3.

The P1 incidence is higher in each specified year than the RS incidence though a moderate treatment-induced infectivity reduction factor is adopted in both simulations. We observe an increases of 10.95% in 2030 and of 2.87% in 2040, respectively.

Once the optimistic therapy is adopted in the following P2-P4 simulations, the resulting incidence shows an obvious overall downtrend compared to the RS incidence, since down arrows are predominantly present in the P2, P3 and P4 columns of Table 3. Among incidence resulting from these 3 simulations, the P2 incidence declines to the largest extent, e.g., by 11.24% in 2030 and 28.7% in 2040.

Surprisingly, for the P5 simulation an overall uptrend occurs compared to the RS incidence, since up arrows are predominantly present in the P5 column of Table 3. Although the optimistic treatment-induced infectivity reduction factor is used for all P2-P5 simulations, the P5 incidence shows a totally different upgoing trend compared to the largely downgoing incidence resulting from the P2, P3 and P4 simulations in the long run. For example, in 2035 there is a 5.5%↑ for P5 incidence, while there are 26.3%↓, 10.7%↓ and 9.63%↓ for P2, P3 and P4 incidence, respectively. Additionally, from 2030 onward the P5 incidence approximately remains more or less at the same level as the RS incidence, for instance, 1.64 versus 1.63 in 2030 and 1.58 versus 1.60 in 2040, respectively.

In addition, the yearly fraction of the diagnosed (the number of the diagnosed divided by the number of the infected), inferred from our simulated results, is fluctuating within the range of 0.35-0.56 and finally approaching 0.41 (data not shown).

Table 3: Resulting incidence from simulations of scenarios. Data points of incidence, taken at 5-year intervals after 2005 from the results of scenarios simulations, are listed and compared. In each year the incidence resulting from every prediction scenario is compared percentagewise to that of the reference scenario (*), respectively.

| Year | Simulated incidence | | | | | |
|---|---|---|---|---|---|---|
| | RS* | P1 | P2 | P3 | P4 | P5 |
| 2010 | 1.61423 | 1.76164 | 1.62148 | 1.7568 | 1.8583 | 2.18211 |
| | - | 9.13% ↑ | 0.45% ↑ | 8.83% ↑ | 15.12% ↑ | 35.18% ↑ |
| 2015 | 1.69397 | 1.77855 | 1.68189 | 1.75922 | 1.87763 | 2.15311 |
| | - | 4.99% ↑ | 0.71% ↓ | 3.85% ↑ | 10.84% ↑ | 27.1% ↑ |
| 2020 | 1.73505 | 1.79063 | 1.50065 | 1.56831 | 1.84138 | 2.0057 |
| | - | 3.2% ↑ | 13.51% ↓ | 9.61% ↓ | 6.13% ↑ | 15.6% ↑ |
| 2025 | 1.60215 | 1.77855 | 1.54657 | 1.57556 | 1.59248 | 1.71331 |
| | - | 11.01% ↑ | 3.47% ↓ | 1.66% ↓ | 0.6% ↓ | 6.94% ↑ |
| 2030 | 1.63356 | 1.81238 | 1.44991 | 1.40399 | 1.52482 | 1.64081 |
| | - | 10.95% ↑ | 11.24% ↓ | 14.05% ↓ | 6.66% ↓ | 0.44% ↑ |
| 2035 | 1.5804 | 1.76647 | 1.16476 | 1.41124 | 1.42816 | 1.66739 |
| | - | 11.77% ↑ | 26.3% ↓ | 10.7% ↓ | 9.63% ↓ | 5.5 % ↑ |
| 2040 | 1.59973 | 1.64564 | 1.14059 | 1.3025 | 1.37016 | 1.57556 |
| | - | 2.87% ↑ | 28.7% ↓ | 18.58% ↓ | 14.35% ↓ | 1.51 % ↓ |

Discussion

We construct an agent-based probabilistic Complex Agent Network model to gain increased insight into the spread of HIV among men who have sex with men in Amsterdam. In our approach, the *in silico* population is mimicked as sexual contact networks and the transmission dynamics of HIV is embedded in those networks describing complex interactions among MSM. Sexual contacts, steady or casual, between each pair of partners that have individual disease status, contribute to virus transmission from the infected to the susceptible. The simulated incidence in 1985-2006 is able to reproduce the ACS historical incidence well (Figure 1 and Figure 2a), based on the null-hypothesis T-test performed. We examine how two factors of vital significance–the treatment-induced infectivity reduction factor and risk behavior factor–influence the HIV spreading by performing a set of scenario simulations. The resulting analyses (Table 3) reveal that in the long term the effects of the optimistic therapy can be counterbalanced by an increase in risk behavior of at least 30% for MSM. This is in line with the conclusion drawn by Wilson et al., that the risk of HIV transmission in male homosexual partnerships is high over repeated exposures even in the presence of effective treatment [33].

Our research takes advantage of the statistical details of MSM's behavior obtained from the ACS, showing a promising way of simulating HIV spreading among a specific group of people. In contrast, many other modeling studies were not able to incorporate this kind of detailed data. This study provides an alternative and reliable approach to model virus spreading in a highly connected population with individual details. The yearly fraction of the diagnosed is relatively low, restricting the positive influence of highly effective therapy. This low value is consistent with what has been observed in reality. For example, Xiridou et al., stated that 42% of the infected people at the asymptomatic stage know they are HIV positive (diagnosed) [20]; Gras et al., concluded that observed annual proportions of diagnoses for homosexual men were decreasing from about 57% in 1996 to 42% in 2004 taking into account all data in 1996-2004, or approximately remaining at 44% taking into account data in 2000-2004 [34]. Accordingly, many infected people are unaware of their infection due to low fraction of diagnosis and thereby will not benefit from highly effective therapy. Extreme preventive measures have been proposed such as universal testing followed by immediate treatment [35]. But this approach is not practical, as annual universal testing is a logistical challenge and immediate treatment regardless of CD4 count is not in line with current treatment guidelines. Policy makers therefore consider much more subtle and realistic countermeasures.

The simulations demonstrates that the incidence varies in different time points (Figure 2). We believe that this sort of rapid fluctuations in incidence is inherent and is likely the result of network structures in combination with the stochastic nature of the models.

More precise data is needed to estimate the infectiousness of people who are receiving therapy, taking into consideration their adherence, behavior changes since receiving therapy and rates of resistance emergence.

Also, a better model of estimating risk behavior is expected to further improve our understanding of the underlying determinants of risk behavior. For example, the fraction of MSM that use condoms without any exception or only during the last insertive anal intercourse can be introduced.

The transmission model has several challenges. Firstly, the model is validated by comparing the simulated yearly incidence to the ACS historical incidence, ignoring prevalence and other indicators. This is mainly due to the availability of the ACS incidence data, and also the the dimensionless feature of the incidence since the number of recruited MSM in the ACS varies each year while the network size remains constant throughout. To avoid a possible bias resulting from this incidence-biased process, we can supplement comparisons according to indicators other than incidence.

Secondly, for practicability, we assume that HAART can to an optimistic extent reduce patients' infectiousness by up to 99% instead of 100%. There exists a serious controversy on the effectiveness of HAART. Although HAART could reduce viral load to an undetectable level, Edwin J Bernard stated that being undetectable does not necessarily mean being entirely uninfectious [36]. Additionally, quantifying the proportion of people who will reach undetectable viral load due to therapy still relies on further clinical surveillance and statistics.

The advantages of using the Complex Agent Network model have not been fully exploited in this study; for example we use neither an individually dynamic risk behavior factor nor consider individual histories of risky contact (due to the lack of data). Incorporating more data-driven details may result in dramatic increases in computational cost to which modeling a population in hierarchical networks can be a promising solution [37].

## Conclusion

Our results suggest that the positive influence of effective therapy can be outweighed by an increase in risk behavior of at least 30% for MSM. This implies that if universal voluntary HIV testing is currently impossible or impractical, lowering risk behavior is the prominent control mechanism of HIV incidence even in the presence of effective treatment. This is an essential result since recent studies indicate that even in the presence of potent antiretroviral therapy, the incidence of HIV among MSM still increases.

## Competing interests

The authors declare that they have no competing interests.

Authors contributions

SM carried out the majority of the model design and construction, and the model implementation, all analyses, interpreted the results and prepared the manuscript as the lead writer. RQ and PMAS participated in the model design and validation, and helped to revise the manuscript. DV participated in the data collection and helped to revise the manuscript. YZ participated in the conception of the complex agent network approach. All authors read and approved the final manuscript.

Acknowledgements

The authors would like to acknowledge the financial support of the severe infectious diseases spreading research based on social networks (Chinese grant 2008ZX10004-013) and the European DynaNets (www.dynanets.org) grant (EU Grant Agreement Number 233847). In particular, we thank Dr. Viktor Müller for his helpful suggestions.

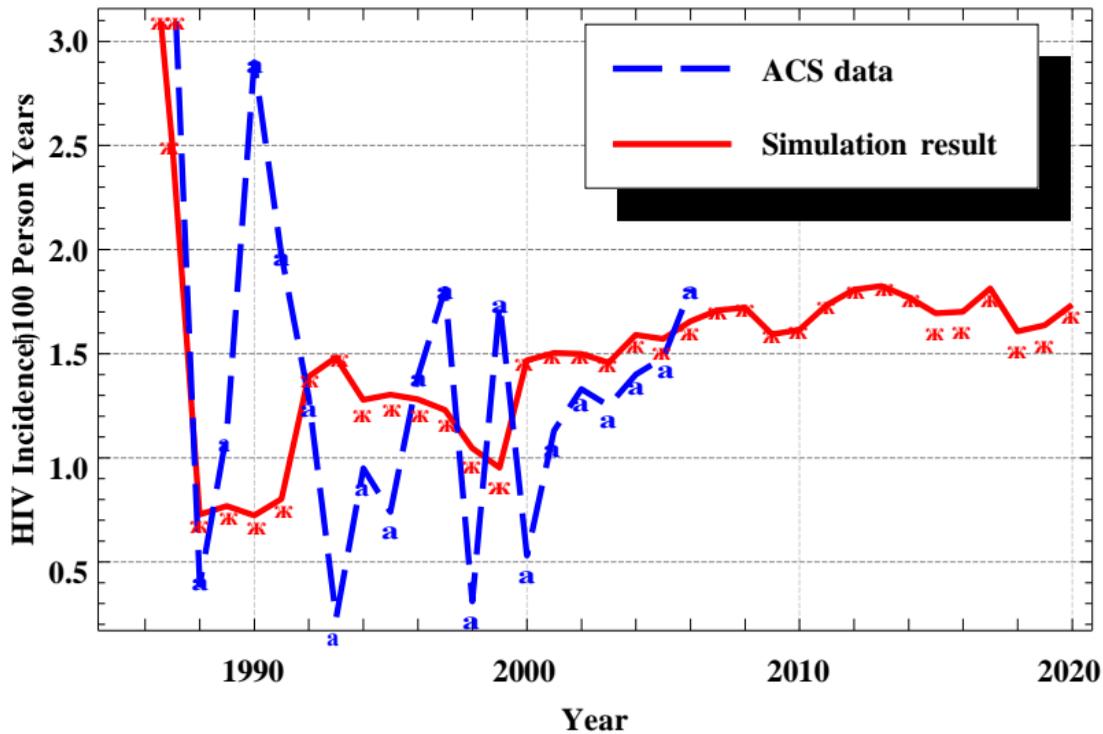

Figure 1

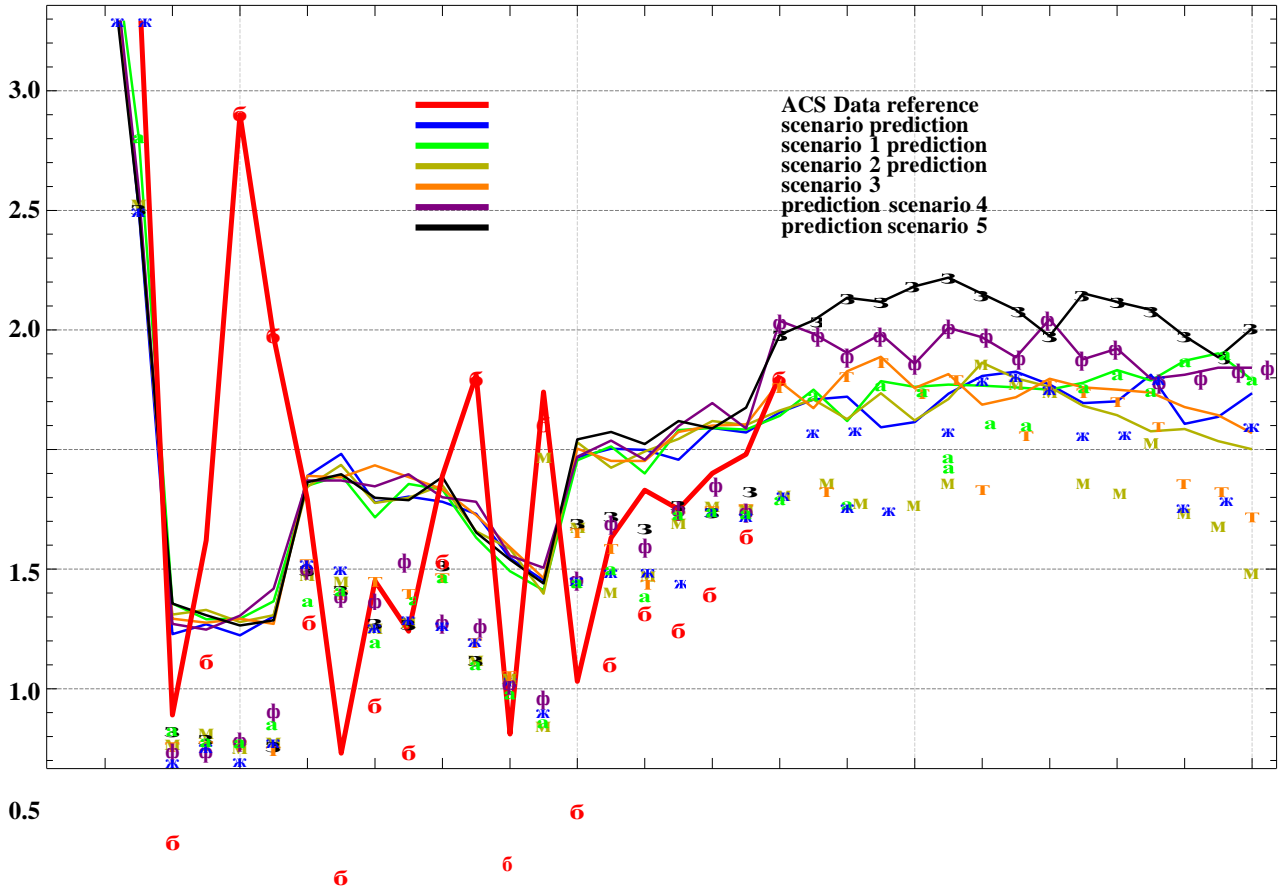

Figure 2

# HIV Incidence/100 Person Years

**Incidence/100 Person Years**

|      |      |      |      |
|------|------|------|------|
| 1990 | 2000 | 2010 | 2020 |

**Year**

Figure 3

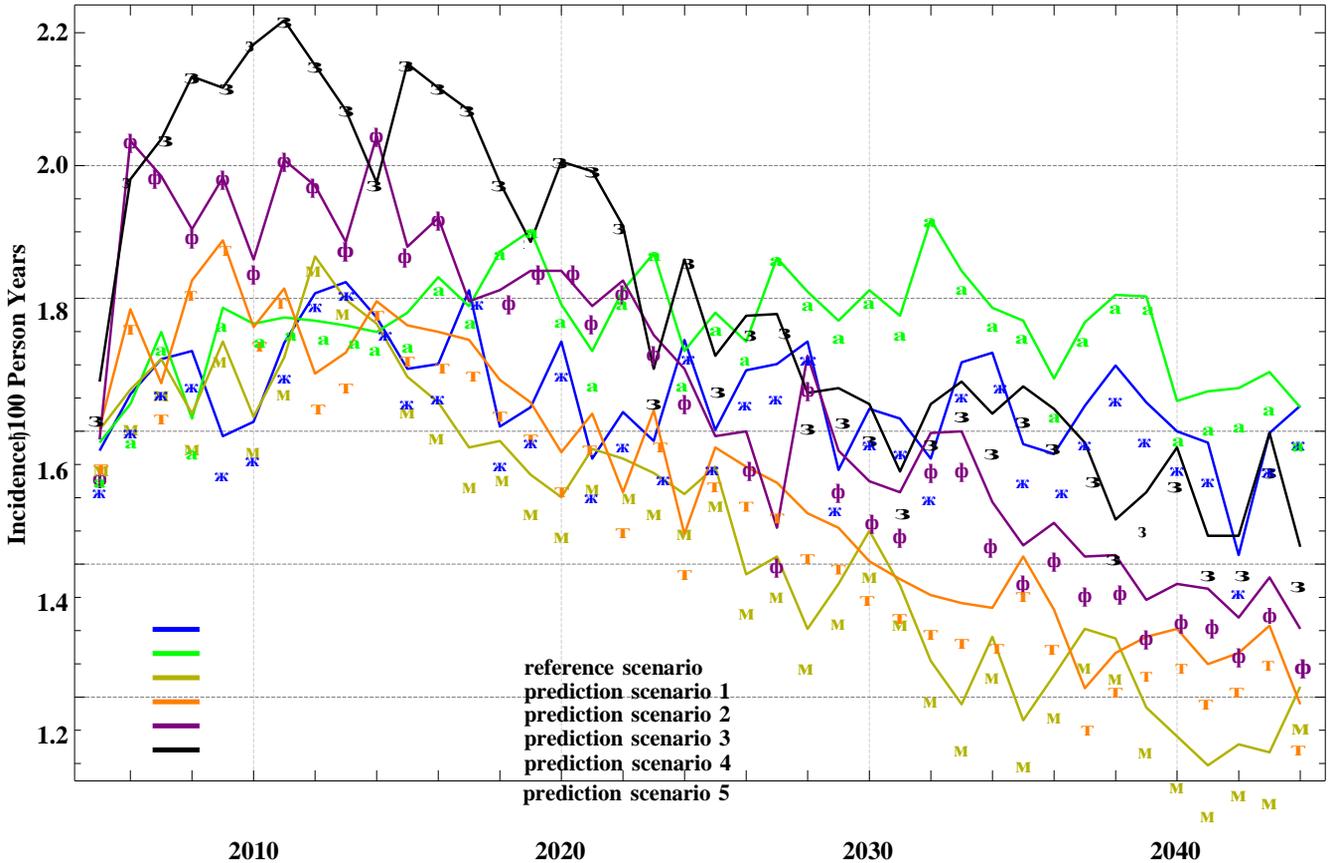

Figure 4

**HIV Incidence/100 Person Years**

Year

Figure 5